\documentclass[prb,twocolumn,showpacs,preprintnumbers,amsmath,amssymb,floatfix]{revtex4}

\usepackage{graphicx}
\usepackage{dcolumn}
\usepackage{bm}

\begin{document}
\bibliographystyle{apsrev}

\title{Electronic charge reconstruction of doped Mott insulators in multilayered nanostructures}

\author{Ling Chen}
\author{J. K. Freericks}%
\email{freericks@physics.georgetown.edu}
\affiliation{Department of Physics, Georgetown University,
Washington, DC 20057, USA }

\date{\today}

\begin{abstract}
Dynamical mean-field theory is employed to calculate the
electronic charge reconstruction of multilayered inhomogeneous
devices composed of semi-infinite metallic lead layers sandwiching
barrier planes of a strongly correlated material (that can be
tuned through the metal-insulator Mott transition). The main focus
is on barriers that are doped Mott insulators, and how the
electronic charge reconstruction can create well-defined Mott
insulating regions in a device whose thickness is governed by
intrinsic materials properties, and hence may be able to be
reproducibly made.
\end{abstract}

\maketitle
\section{\normalsize{Introduction}}
Understanding the interface properties of strongly correlated
electron systems placed into inhomogeneous environments on the
nanoscale combines the fields of strongly correlated electron
systems and nanotechnology. The interface properties of strongly
correlated systems will play an important role in determining the
properties of devices made from these materials. One of the
important interface properties is electronic charge
reconstruction~\cite{millis}. At nearly all types of
metal-semiconductor interfaces, a so called Schottky barrier
exhibits charge depletion in the doped semiconductor region close
to the interface. Because the Fermi energy in the metal differs
from that in the semiconductor, mobile carriers in the
semiconductor side of the barrier diffuse into the metal side
until a static equilibrium is reached. Recent experiments
conducted on strongly correlated materials have shown a similar
interface-induced charge reconstruction in inhomogeneous
nanostructures. Ohtomo and co-workers~\cite{Ohtomo} have
fabricated atomically precise digital heterostructures consisting
of a controllable number of planes of LaTiO$_3$ (a correlated
electron Mott-insulating material) separated by a controllable
number of planes of SrTiO$_3$ (a more conventional band-insulating
material). The experiment demonstrated an insulator-metal
crossover near the interface due to electronic charge
reconstruction. The insulating heterostructure developed
conducting channels near the interfaces for current parallel to
the planes. Okamoto and Millis~\cite{millis} found, through a
Hartree-Fock calculation, that the electronic charge
reconstruction leads to metallic behavior at the interface between
the two insulators. This is because the mismatch of chemical
potentials creates screened dipole interface charge
reconstructions that are conducting due to the excess or deficit
of charge. Varela and collaborators~\cite{Varela} have provided
evidence for extensive charge reconstruction from
manganite-cuprates heterostructures at a
YBa$_{2}$Cu$_3$O$_{7-x}$/La$_{0.67}$Ca$_{0.33}$MnO$_3$ interface,
which also exhibits a metal-insulator crossover near the
interface. From a theoretical standpoint, Nikoli$\acute{c}$,
Freericks and Miller~\cite{Freericks} developed a semiclassical
approach to the Potthoff-Nolting~\cite{Potthoff} algorithm to
create a dynamical mean-field theory description of electronic
charge reconstruction and applied it to
superconductor-insulator-normal metal-insulator-superconductor
Josephson junctions.

In this work, we investigate the electronic charge reconstruction
of a doped Mott-insulator, where the reconstruction can dope parts
of the system close to the insulating phase. Such a system may be
realized in high temperature superconducting grain boundaries,
where the electrically active grain boundary can lead to an
electronic charge reconstruction~\cite{mannhart}. We take a
semi-infinite ballistic-metal lead and couple it to another
semi-infinite ballistic-metal lead through a strongly correlated
barrier material of varying thickness. By adjusting the
interaction and the filling of the barrier material, we study
interface properties for different types of materials. Here we
emphasize the physics of the doped Mott-insulator.

The organization of this paper is as follows: in Sec. II, we
present a detailed derivation of the formalism and the numerical
algorithms used to calculate the charge reconstruction of
nanostructures. In Sec. III, we describe our numerical results. We
end with our conclusions in Sec. IV.

\section{\normalsize{Formalism}}
We apply the Potthoff-Nolting~\cite{Potthoff} approach to
multilayered nanostructures, which involve translationally
invariant $x-y$ planes stacked in the longitudinal $z$-direction.
We choose square lattice planes stacked along $z$ direction, so
the lattice sites are identical to those of a simple cubic
lattice. We have periodicity in the $x$ and $y$ directions, but we
allow inhomogeneity in the $z$-direction. All interactions are
also translationally invariant within each plane, but can change
from one plane to the next. The system is described by a spatially
inhomogeneous many-body problem. Potthoff and Nolting described
the idea of using a mixed basis for inhomogeneous DMFT: First,
Fourier transform the $x$ and $y$ coordinates to wavevectors $k_x$
and $k_y$ but keep the $z$-component in real space. Then for each
two-dimensional band energy, we have a quasi-one-dimensional
problem to solve, which has a tridiagonal representation in real
space, and can be solved with the so called quantum zipper
algorithm~\cite{zipper}.

Because of the translational invariance in each two-dimensional
plane, we can describe the intraplane hopping via a
two-dimensional bandstructure, which becomes

\begin{equation}
\epsilon^{\|}_{\alpha}(\mathbf{k}_{x},\mathbf{k}_{y})=-2t^{\|}_{\alpha}[\cos(\mathbf{k}_{x})+\cos(\mathbf{k}_{y})],
\label{2}
\end{equation}
for a square lattice plane, where $t^{\|}_{\alpha}$ is the
nearest-neighbor intraplane hopping on the $\alpha$th plane and
($k_x,k_y,0$) is the two-dimensional wave vector.

For the interaction, we employ the Falicov-Kimball
model~\cite{Falicov} which involves an interaction between
conduction electrons with localized particles
(\textit{f}-electrons or charged ions) when the conduction
electron hops onto a site occupied by the localized particle. The
Falicov-Kimball model has a non-Fermi liquid ground state in the
metallic regime, because the electrons see static charge
scatterers, which always produce a finite scattering lifetime, so
there is no quasiparticle resonance, as seen in other strongly
correlated models like the Hubbard~\cite{Hubbard} and periodic
Anderson~\cite{Anderson} model. It also has a Mott-type
metal-insulator transition, that sets in when the correlation
strength is large enough and the total number of the particles
(localized plus delocalized) equals the number of lattice sites.
This is because the energy cost for double occupation becomes too
high if $U$ is large enough, and the system becomes an insulator;
the metal insulator transition can occur for systems without
particle-hole symmetry, unlike the single-band Hubbard model which
always has the Mott transition precisely at half
filling~\cite{Demchencko}. Our choice of using the Falicov-Kimball
model is pragmatic, since the DMFT can be easily solved for this
system and it describes interesting metal-insulator transitions of
strongly correlated materials. We expect the results in the
insulating phase to resemble other correlated insulators, since
the most important property of a correlated insulator is the size
of its gap. On the metallic side, the Falicov-Kimball model is
good for describing the crossover from ballistic to diffusive
transport in dirty metals, but it is unable to describe the
coherent quasiparticle formation, with a renormalized Fermi
energy, seen in pure models like the Hubbard model. We feel it is
nevertheless an interesting model to consider for examining
systems near a Mott transition (especially since any experimental
system will always have disorder, so the renormalized Fermi liquid
will also disappear close to the Mott transition due to this
disorder, and the actual metal-insulator transition may be closer
to the scenario of the Falicov-Kimball model). We consider
spinless electrons here, but spin can be included by introducing a
factor of 2 into some of the results; but note that it will modify
the self-consistency relation for the Coulomb potential energy,
which will not involve just a trivial change of the final
converged results (the Coulomb potential energy will be doubled in
magnitude, but that doubled value goes into the local chemical
potential for the next iteration of the algorithm). Introducing
spin also changes the filling condition for the Mott-transition.

The charge reconstruction at the interface leaves net charge on
each plane. So the Hamiltonian needs to be altered to include this
effect. We analyze this problem using a semiclassical
approach---the charge rearrangements are employed to determine
classical electric fields, electric potentials and potential
energies, which are then input into the Hamiltonian, that is
subsequently solved using quantum mechanics. We assume that the
electric charge is uniformly spaced over the plane. Then the
electric field is a constant, perpendicular to the plane. If the
net charge density on plane $\alpha$ is
$\rho_{\alpha}-\rho_{\alpha}^{bulk}$ and the permittivity is
$\epsilon_{\alpha}$, then the change in polarization at the
interface between the two dielectric planes induces an additional
polarization charge on the interface that leads to a discontinuous
jump in the electric field halfway between the two lattice planes.
The local field created by the $\alpha$th plane has magnitude
\begin{equation}
|E|=\frac{ea(\rho_\alpha-\rho^{bulk}_{\alpha})}{2\epsilon_{\alpha}},
\label{22}
\end{equation}
in which $a$ is the lattice spacing along $z$ direction. Once the
total fields at each plane are known, we integrate them to obtain
the electric potentials. The discontinuity in the electric field
occurs at the midpoint between the two lattice planes when there
is a change in permittivity.

\begin{figure}
\centering
\includegraphics[angle=-90,scale=0.3,clip=]{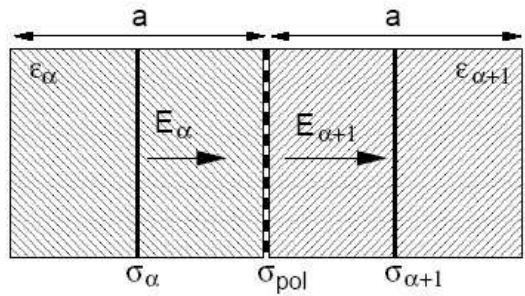}
\caption{Geometry taken for the classical electrostatics problem.
We show a blow up of two planes, $\alpha$ and $\alpha+1$. Assuming
that the excess surface charge density on plane $\alpha$ is
$(\rho_{\alpha}-\rho^{bulk}_{\alpha})a = \sigma_{\alpha}$ and that
the permittivity is $\varepsilon_{\alpha}$ (and similarly for the
$\alpha+1$ plane), then the change in polarization at the
interface between the two dielectric planes induces a polarization
charge on the interface (denoted $\sigma_{pol}$) that leads to a
discontinuous jump in the electric field halfway between the two
lattice planes. Once all the fields are known, we integrate the
total field to get the electric potential. Note that the
discontinuity in the electric field occurs at the midpoint between
the two lattice planes.}
\end{figure}

Performing the integration yields the potential energy due to the
Coulomb interaction of the electronic charge reconstruction as
\begin{widetext}
\begin{equation}
V_{\beta}=-\sum_\alpha(\rho_\alpha-\rho^{bulk}_{\alpha})\times
\left \{ \begin{array}{lc}
  \sum_{\gamma=\alpha+1}^{\beta}\frac{1}{2}[e_{Schot}(\gamma)+e_{Schot}(\gamma-1)],&\beta>\alpha \\
  0,&\beta=\alpha \\
  \sum_{\gamma=\beta+1}^{\alpha}\frac{1}{2}[e_{Schot}(\gamma)+e_{Schot}(\gamma-1)],
&\beta<\alpha
\end{array}\right.
\label{23}
\end{equation}
\end{widetext}
where we define the symbol $e_{Schot}(\alpha)=e^2
a/2\epsilon_{\alpha}$, which is related to the screening length in
a particular medium. This parameter has the units of an energy
multiplied by an area; the product of $e_{Schot}$ with the local
DOS has the units of the inverse of a length, and this is what
determines the decay length of the charge profile. In our
calculations we will set $e_{Schot}$ to a particular value to fix
the screening length.

The Coulomb potential
energies are input into the Hamiltonian as follows: they are
treated by shifting the chemical potential
$\mu\longrightarrow\mu-V_{\alpha}$ on each plane, depending on
what the Coulomb potential energy is. The Falicov-Kimball
Hamiltonian with electronic charge reconstruction is then
\begin{eqnarray}
H&=&-\sum_{\alpha i\beta j}t_{\alpha i\beta j}c^{\dag}_{\alpha
i}c^{}_{\beta j}+\sum_{\alpha i}U_{\alpha i}c^{\dag}_{\alpha
i}c^{}_{\alpha i}f^{\dag}_{\alpha i}f^{}_{\alpha
i}\nonumber
\\&-&\sum_{\alpha i} (\mu+\Delta E_{F\alpha}-V_{\alpha}) c^{\dag}_{\alpha i}
c^{}_{\alpha i}, \label{1}
\end{eqnarray}
in which $\alpha$ and $\beta$ represent different planes and $i$
and $j$ represent lattice sites on the respective planes. The
chemical potential $\mu$ is fixed by the bulk value of the leads.
$\Delta E_{F\alpha}$ is the mismatch of the center of the bands
between the leads and the barrier ($\Delta E_{F\alpha}=0$ in the
leads). By adjusting the value of $\Delta E_{F\alpha}$, we control
the relative shift of the bands inside the multilayered
nanostructure; indeed, the electronic charge reconstruction occurs
because $\Delta E_f$ is adjusted to describe the chemical
potential mismatch between the two materials at some given
temperature. Note that $\Delta E_f$ is a fixed constant that does
not change with the temperature.

To analyze the many-body problem, we use a Green's function
technique. The imaginary-time Green's function, in real space, is
defined by
\begin{equation}
G_{\alpha i\beta j}(\tau)=-\langle T_{\tau} c_{\alpha
i}(\tau)c^{\dag}_{\beta j}(0) \rangle. \label{3}
\end{equation}
for imaginary time $\tau$. The notation $\langle \mathcal{O}
\rangle$ denotes ${\rm Tr} T_{\tau} \exp(-\beta[H-\mu N])
\mathcal{O}/\tilde{Z}$, where $\tilde{Z}$ is the partition
function $\tilde{Z}={\rm Tr} \exp(-\beta[H-\mu N])$. The operators are
expressed in the Heisenberg representation where
$\mathcal{O}(\tau)=\exp(\tau[H-\mu N])\mathcal{O}\exp(-\tau[H-\mu
N])$. The symbol $T_{\tau}$ denotes time ordering of the
operators, with earlier $\tau$ values appearing to the right and
$\beta$ is the inverse temperature, $\beta=1/T$. We will work with
the Matsubara frequency Green's functions, defined for imaginary
frequencies $i\omega_n=i \pi T(2n+1)$. These are determined by a
Fourier transformation
\begin{equation}
G_{\alpha i \beta j}(i\omega_n)=\int^{\beta}_{0} d\tau
e^{i\omega_{n}\tau} G_{\alpha i\beta j}(\tau). \label{4}
\end{equation}
We can write the equation of motion for the Green's function in
real space, which satisfies
\begin{eqnarray}
G^{-1}_{\alpha i\beta j}(i\omega_n)&=&(i\omega_n+\mu+\Delta
E_{F\alpha}-V_{\alpha})\delta_{\alpha i\beta
j}\nonumber\\&-&\Sigma_{\alpha}(i\omega_{n})\delta_{\alpha i\beta
j}+t_{\alpha i\beta j}. \label{5}
\end{eqnarray}
Now we go to a mixed basis, by Fourier transforming in the $x$ and
$y$ directions, to find
\begin{eqnarray}
G^{-1}_{\alpha
\beta}(\mathbf{k},i\omega_{n})&=&[i\omega_{n}+(\mu+\Delta
E_{F\alpha}-V_{\alpha})-\Sigma_{\alpha}(i\omega_{n})\label{6}\\
&-&\epsilon^{\|}_{\alpha}(\mathbf{k})]\delta_{\alpha
\beta} + t_{\alpha \alpha+1}\delta_{\alpha+1 \beta}+t_{\alpha
\alpha-1}\delta_{\alpha-1 \beta}, \nonumber
\end{eqnarray}
with $\Sigma_{\alpha}(i\omega_{n})$ the local self-energy for
plane $\alpha$ and $\mathbf{k}$ the two-dimensional planar
momentum. Finally, we use the identity
$\sum_{\gamma}G_{\alpha\gamma}({\bf k})G^{-1}_{\gamma\beta}({\bf k})
=\delta_{\alpha\beta}$ to arrive at the starting point for the recursive solution
to the problem,
\begin{eqnarray}
\delta_{\alpha \beta}&=&G_{\alpha
\beta}(\mathbf{k},i\omega_{n})\nonumber\\
&\times&[i\omega_{n}+(\mu+\Delta
E_{F\beta}-V_{\beta})-\Sigma_{\beta}(i\omega_{n})-\epsilon^{\|}_{\beta}(\mathbf{k})]
\nonumber\\
&+&G_{\alpha \beta-1}(\mathbf{k},i\omega_{n})t_{\beta-1
\beta}+G_{\alpha \beta+1}(\mathbf{k},i\omega_{n})t_{\beta+1
\beta}. \label{7}
\end{eqnarray}
It turns out that there is a straightforward procedure to
determine $G_{\alpha \beta}$ from this equation of motion. It is
called the quantum zipper algorithm. We start with $\beta=\alpha$,
which can be used to find the Green's function via
\begin{widetext}
\begin{equation}
G_{\alpha
\alpha}(\mathbf{k},i\omega_{n})=\frac{1}{i\omega_{n}+(\mu+\Delta
E_{F\alpha}-V_{\alpha})-\Sigma_{\alpha}(i\omega_{n})-\epsilon^{\|}_{\alpha}(\mathbf{k})+\frac{G_{\alpha
\alpha-1}(\mathbf{k},i\omega_{n})}{G_{\alpha
\alpha}(\mathbf{k},i\omega_{n})} t_{\alpha-1
\alpha}+\frac{G_{\alpha
\alpha+1}(\mathbf{k},i\omega_{n})}{G_{\alpha
\alpha}(\mathbf{k},i\omega_{n})} t_{\alpha+1 \alpha}}. \label{8}
\end{equation}
Next, we consider the equations with $\beta\neq\alpha$, which can
be put into the form
\begin{equation}
-\frac{G_{\alpha \alpha-m+1}(\mathbf{k},i\omega_{n})t_{\alpha-m+1
\alpha-m}}{G_{\alpha
\alpha-m}(\mathbf{k},i\omega_{n})}=i\omega_{n}+(\mu+\Delta
E_{F\alpha-m}-V_{\alpha-m})-\Sigma_{\alpha-m}(i\omega_{n})-\epsilon^{\|}_{\alpha-m}(\mathbf{k})
+\frac{G_{\alpha \alpha-m-1}(\mathbf{k},i\omega_{n})t_{\alpha-m-1
\alpha-m}}{G_{\alpha \alpha-m}(\mathbf{k},i\omega_{n})}, \label{9}
\end{equation}
\end{widetext}
for $m > 0$, with a similar result for the recurrence with $m<0$.
In these equations, we have used $\Sigma_\alpha$ to denote the
local self-energy $\Sigma_{\alpha\alpha}$ on plane $\alpha$. We
define the left function
\begin{equation}
L_{\alpha-m}(\mathbf{k},i\omega_{n})=-\frac{G_{\alpha
\alpha-m+1}(\mathbf{k},i\omega_{n})t_{\alpha-m+1
\alpha-m}}{G_{\alpha \alpha-m}(\mathbf{k},i\omega_{n})} \label{10}
\end{equation}
and then determine the recurrence relation from Eq.~(\ref{9})
\begin{eqnarray}
L_{\alpha-m}(\mathbf{k},i\omega_{n})=i\omega_{n}+(\mu+\Delta
E_{F\alpha-m}-V_{\alpha-m})-\nonumber\\\Sigma_{\alpha-m}(i\omega_{n})-\epsilon^{\|}_{\alpha-m}(\mathbf{k})
-\frac{t_{\alpha-m \alpha-m-1}t_{\alpha-m-1
\alpha-m}}{L_{\alpha-m-1}(\mathbf{k},i\omega_{n})}. \label{11}
\end{eqnarray}
We solve the recurrence relation by starting with the result for
$L_{-\infty}$, and then iterating Eq.~(\ref{11}) up to $m = 0$.
Since we must have a finite set of equations for an actual calculation, we
assume we have semi-infinite metallic leads, hence we can
determine $L_{-\infty}$ by substituting $L_{-\infty}$ into both
the left and right hand sides of Eq.~(\ref{11}) with $\Delta
E_{F\alpha}+V_{\alpha}=0$, which produces a quadratic equation for
$L_{-\infty}$ that is solved by
\begin{eqnarray}
&~&L_{-\infty}(\mathbf{k},i\omega_{n})=\frac{i\omega_{n}+\mu-\Sigma_{-\infty}(i\omega_{n})-\epsilon^{\|}_{-\infty}(\mathbf{k})}{2}\nonumber\\
&\pm&\frac{\sqrt{[i\omega_{n}+\mu-\Sigma_{-\infty}(i\omega_{n})-\epsilon^{\|}_{-\infty}(\mathbf{k})]^{2}-4t^{2}_{-\infty}}}{2}.
\label{12}
\end{eqnarray}
This determines the left functions far from the interface. The
sign in Eq.~(\ref{12}) is chosen to yield an imaginary part less
than zero for $i\omega_{n}$ lying in the upper half plane, and
vice versa for $i\omega_{n}$ lying in the lower half plane. When
we are sufficiently far from the interface, the Green's functions
will be essentially the same as the bulk and hence the
$L_{\alpha}$ functions will equal $L_{-\infty}$. In our
calculations we allow $L_\alpha$ to differ from $L_{-\infty}$ only
for the thirty planes closest to the interface on either side of
the barrier. This means we start the recurrence relation in
Eq.~(\ref{11}) with $\alpha-m-1$ being the thirty first plane to
the left. Then all subsequent $L_{\alpha}$'s are allowed to vary
until we reach 31 planes to the right of the barrier, where we
assume $L_{\alpha}$ becomes a constant again. This approach is
accurate, when the system heals to its bulk values within those
thirty planes on either side of the interface. If this healing has
not occurred, then one needs to include more planes before one
terminates the problem with the semi-infinite bulk solution.

In a similar fashion, we define a right function and a recurrence
relation to the right, with the right function being
\begin{equation}
R_{\alpha+m}(\mathbf{k},i\omega_{n})=-\frac{G_{\alpha
\alpha+m-1}(\mathbf{k},i\omega_{n})t_{\alpha+m-1
\alpha+m}}{G_{\alpha \alpha+m}(\mathbf{k},i\omega_{n})} \label{13}
\end{equation}
and the recurrence relation satisfying
\begin{eqnarray}
R_{\alpha+m}(\mathbf{k},i\omega_{n})=i\omega_{n}+(\mu+\Delta
E_{F\alpha+m}-V_{\alpha+m})-\nonumber\\
\Sigma_{\alpha+m}(i\omega_{n})-\epsilon^{\|}_{\alpha+m}(\mathbf{k})
-\frac{t_{\alpha+m \alpha+m+1}t_{\alpha+m+1
\alpha+m}}{R_{\alpha+m+1}(\mathbf{k},i\omega_{n})}. \label{14}
\end{eqnarray}
We solve the right recurrence relation by starting with the result
for $R_{\infty}$, and then iterating Eq.~(\ref{14}) up to $m = 0$.
As before, we determine $R_{\infty}$ by substituting $R_{\infty}$
into both the left and right hand sides of Eq.~(\ref{14}), where
$\Delta E_{F\alpha}-V_{\alpha}=0$,
\begin{eqnarray}
&~&R_{\infty}(\mathbf{k},i\omega_{n})=\frac{i\omega_{n}+\mu-\Sigma_{\infty}(i\omega_{n})-\epsilon^{\|}_{\infty}(\mathbf{k})}{2}\nonumber\\
&\pm&\frac{\sqrt{[i\omega_{n}+\mu-\Sigma_{\infty}(i\omega_{n})-\epsilon^{\|}_{\infty}(\mathbf{k})]^{2}-4t^{2}_{\infty}}}{2}.
\label{15}
\end{eqnarray}
The sign in Eq.~(\ref{15}) is chosen the same way as in Eq.~(\ref{12}).
In our calculations, we also assume that the right
function is equal to the value $R_{\infty}$ found in the bulk,
until we are within thirty planes of the first interface. Then we
allow those thirty planes to be self-consistently determined with
$R_{\alpha}$ possibly changing, and we include a similar thirty
planes on the left hand side of the last interface, terminating
with the bulk result to the left as well. Using the left and right
functions, we finally obtain the Green's function from
\begin{widetext}
\begin{equation}
G_{\alpha\alpha}(\mathbf{k},i\omega_{n})=
\frac{1}{L_{\alpha}(\mathbf{k},i\omega_{n})+R_{\alpha}(\mathbf{k},i\omega_{n})-[i\omega_{n}+(\mu+\Delta
E_{F\alpha}-V_{\alpha})-\Sigma_{\alpha}(i\omega_{n})-\epsilon^{\|}_{\alpha}(\mathbf{k})]},
\label{16}
\end{equation}
\end{widetext}
where we used Eq.~(\ref{11}) and Eq.~(\ref{14}) in Eq.~(\ref{8}).
This technique for determining the mixed-basis Green's function is
called the quantum zipper algorithm~\cite{zipper}; it has been
modified here to treat the electronic charge reconstruction.

The local Green's function on each plane is then found by summing
over the two-dimensional momenta, which can be replaced by an
integral over the two-dimensional density of states (DOS) since
all the momentum dependence in the algorithm is in terms of
$\epsilon^{\|}_{\alpha}$:
\begin{equation}
G_{\alpha \alpha}(i\omega_{n})=\int d\epsilon^{\|}_{\alpha}
\rho^{2d}(\epsilon^{\|}_{\alpha})G_{\alpha
\alpha}(\epsilon^{\|}_{\alpha},i\omega_{n}), \label{17}
\end{equation}

with
\begin{equation}
\rho^{2d}(\epsilon^{\|}_{\alpha})=\frac{1}{2\pi^{2}t^{\|}_{\alpha}a^{2}}K\left(
1-\sqrt{1-\frac{(\epsilon^{\|}_{\alpha})^{2}}{(4t^{\|}_{\alpha})^{2}}}\right),
\label{18}
\end{equation}
and $K(x)$ the complete elliptic integral of the first kind.

The DMFT algorithm starts with a self-energy on each plane, which
is usually chosen to be zero. Next, we use the quantum zipper
algorithm to find the local Green's function on each plane. This
step is the inhomogeneous nanostructure equivalent to the Hilbert
transform, which is used in bulk DMFT. Once the local Green's
function is known on each plane, we extract the local effective
medium via
\begin{equation}
G_{0\alpha}(i\omega_{n})^{-1}=G_{\alpha}(i\omega_{n})^{-1}+\Sigma_{\alpha}(i\omega_{n}),
\label{20}
\end{equation}
for each plane. Next, we need to solve the local impurity problem
for the given Hamiltonian on the $\alpha$th plane with the given
effective medium:
\begin{equation}
G_{\alpha}(i\omega_{n})=(1-w_{1})G_{0\alpha}(i\omega_{n})+\frac{w_{1}}{G_{0\alpha}(i\omega_{n})^{-1}-U_{\alpha}},
\label{impurity}
\end{equation}
with $w_{1}$ the $f$-electron density. This will produce a new
local Green's function for each plane, and then a new self-energy
for each plane via Eq.~(\ref{20}).

The electronic charge on each plane is calculated by summing the
Green's functions over all Fermionic Matsubara frequencies on the imaginary axis,
\begin{equation}
\rho_{\alpha}=\frac{1}{2}+T\sum_{n}G_\alpha (i\omega_n).
\label{21}
\end{equation}
Since $G_{\alpha}(i\omega_n)$ behaves like ${1}/{i\omega_n}$ at
large $n$, we regularize the summation by subtracting $T\sum_n
1/[i\omega_n-\mu+{\rm Re}\Sigma_{\alpha}(i\omega_{n_{\rm max}})]$
from $G_\alpha (i\omega_n)$ to speed up the convergence. ${\rm
Re}\Sigma_{\alpha}(i\omega_{n_{\rm max}})$ is the real part of the
self energy at the highest Matsubara frequency ($n_{\rm max}$) we
use in our calculation. Typically $n_{\rm max}$ satisfies $2\pi T
n_{\rm max}=30$. Since $T\sum_n\frac{1}{i\omega_n-\mu+{\rm
Re}\Sigma_{\alpha}(i\omega_{n_{\rm max}})}=\tanh\{\beta[\mu-{\rm
Re}\Sigma_{\alpha}(i\omega_{n_{\rm max}})]/2\}/2$, Eq.~(\ref{21})
becomes,
\begin{eqnarray}
\rho_{\alpha}&=&\frac{1}{2}+T\sum_{n}\left[G_\alpha
(i\omega_n)-\frac{1}{i\omega_n-\mu+{\rm
Re}\Sigma_{\alpha}(i\omega_{n_{\rm max}})}\right]
\nonumber\\
&+&\frac{1}{2}\tanh\left ( \frac{\beta[\mu-{\rm
Re}\Sigma_{\alpha}(i\omega_{n_{\rm max}})]}{2}\right ). \label{27}
\end{eqnarray}
Note that our regularization scheme requires that
$T\sum_{n}G_\alpha (i\omega_n)=\langle
c^{\dag}_{\alpha}(0)c_{\alpha}(0)-c_{\alpha}(0)c^{\dag}_{\alpha}(0)
\rangle/2$. Once the change in the charge density is known, we can
calculate the electrical potential $V_{\alpha}$ through
Eq.~(\ref{23}). This is then added to the chemical potential to
determine the electrochemical potential at each plane. Then we
iterate the DMFT algorithm with the new electrochemical potential
until the self-energy at every plane converges and the potentials
no longer change. The iterative equations need to be solved with
an averaging procedure to ensure stability of the solutions. If
the potentials are updated too quickly, then the system of
equations migrates to an unphysical fixed point, or limit cycle.
Typically, we use 0.99 of the old potential (or more) and 0.01 (or
less) of the new potential in each averaging step. It usually
takes more than a thousand iterations to reach the desired
convergence, but because our impurity solver is for the
Falicov-Kimball model, the calculations can still be completed
rapidly.

\section{\normalsize{NUMERICAL RESULTS}}
Previous work on the multilayered nanostructure at half filling
has shown interesting properties~\cite{Freericks}. For example, if
the barrier is metallic, the net charges
($\rho_{\alpha}-\rho^{bulk}_{\alpha}$) on every plane divided by
the amount of the band shift ($\triangle E_{f}$) exhibit scaling:
the curves of the charge reconstruction with different band shifts
essentially lay on top of each other with deviations only
occurring very close to the interface. While this effect is most
likely due to the flatness of the simple cubic lattice density of
states near the band center, the result was still rather striking
in the data.

In this contribution, our focus is on the doped Mott-insulating
phase. The doped Mott-insulator-metal interface is in many ways
similar to the Schottky barrier. The barriers are both insulating;
the doped carriers move to the interfaces as the result of the
band mismatch, causing depletion of carriers close to the
interface inside the barriers. However, the doped Mott-insulator
is of particular interest because the band gap and the small
amount of doped charges generate interesting physics that may be
useful for designing future devices~\cite{Macdonald}.

We reduce the number of parameters in our calculations by assuming
all of the hopping matrix elements are equal to $t$ for nearest
neighbors. While not required,  it allows us to reduce the number
of parameters that we vary in our calculations so that we can
focus on the physical properties with fewer calculations. The
hopping scale $t$ is used as our energy scale. As described above,
we include $30$ self consistent planes in the metallic leads to
the left and to the right of our barrier, which is varied between
$15$ and $30$ planes in our calculations. The screening length, as
determined by the parameter $e_{Schot}(\alpha)$, is about $2.2$
lattice spacings. We choose $e_{Schot}(\alpha)=0.4$ throughout the
device (in both the leads and the barrier). The band shift
$\triangle E_{f}$ can be divided into two parts, $E_{f}(T)$ and
$\mu^{bulk}_{barrier}(T)$. In the following analysis, we plot our
result with respect to $E_{f}(T)$, which is $\triangle
E_{f}-\mu^{bulk}_{barrier}(T)$. The reason for doing this is that if
we shift the band of the barrier to the amount of the bulk
chemical potential of the barrier, the chemical potentials of the
lead and the barrier should be automatically aligned with each
other. Shifting the band less than that ($E_{f}<0$), negative
charges should be attracted to the barrier in order to reach
equilibrium; more than that ($E_{f}>0$), positive charges should
be attracted to the barrier. (Note that because the local Green's functions 
change near the interface, even if $E_f(T)=0$, there will always be a small 
charge reconstruction in the general case.) At half filling, the curves are
particle-hole symmetric between positive and negative $E_{f}$. At
half filling, the bulk chemical potential is independent of the
temperature, so changing the temperature will not have effects on
the mismatch of the bands. At fillings other than half filling,
the chemical potential is dependent on temperature. So changing
temperature can have a similar effect as changing $E_{f}$. When
the temperature is low, the effect of different temperatures on
the charge reconstruction is small because of the large band gap
and also because the change in the bulk chemical potential is
small at low temperatures. In our calculation, we set $T=0.25$. If
we take a reasonable energy scale for our system, such as a
noninteracting bandwidth of 3~eV, then $t=0.25$~eV, and the
temperature corresponds to 750~K.

\begin{figure}
\centering
\includegraphics[angle=0,width=8.5cm,clip=]{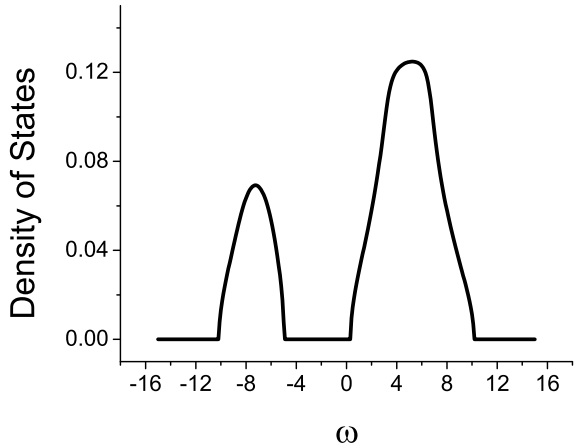}
\caption{Density of states of the central plane in the barrier
versus frequency for $U=12$, $w_1=0.75$, $E_f=0$ and
$\rho_{e}=0.251$. The temperature is $T=0.25$.
With strong interaction, the density of states
has a large gap between the bands. Note the size of the band gap
is roughly 5.0 and if the carriers are drawn from the band across
the gap, $|E_{F}|$ has to be larger than the size of the band
gap. This DOS is indistinguishable from the bulk DOS on this linear-scale
plot. The density of states does not change with $T$, except for the location
of the zero on the frequency axis, due to the temperature dependence of the
chemical potential (as $T\rightarrow 0$, the origin lies just slightly above the
lower band edge of the upper Hubbard band).} \label{dosofmott}
\end{figure}
We tune the Falicov-Kimball interaction in the barrier planes to
$U=12$, which lies well into the Mott-insulating regime. We set
the conduction electron filling at $0.251$ and the density of the
$f$-electron $w_1$ at $0.75$. Hence, we are considering a slightly
doped particle-hole asymmetric Mott insulator. The bulk DOS of the
barrier planes is shown in Fig.~\ref{dosofmott}. The integration
of the DOS over frequency in the lower band is 0.25 and the
integration of the DOS in the upper band is 0.75. Because of the
strong interaction, there is a gap with a size of roughly 5
between the lower band and the upper band.

\begin{figure}
\centering
\includegraphics[angle=0,width=8.5cm,clip=]{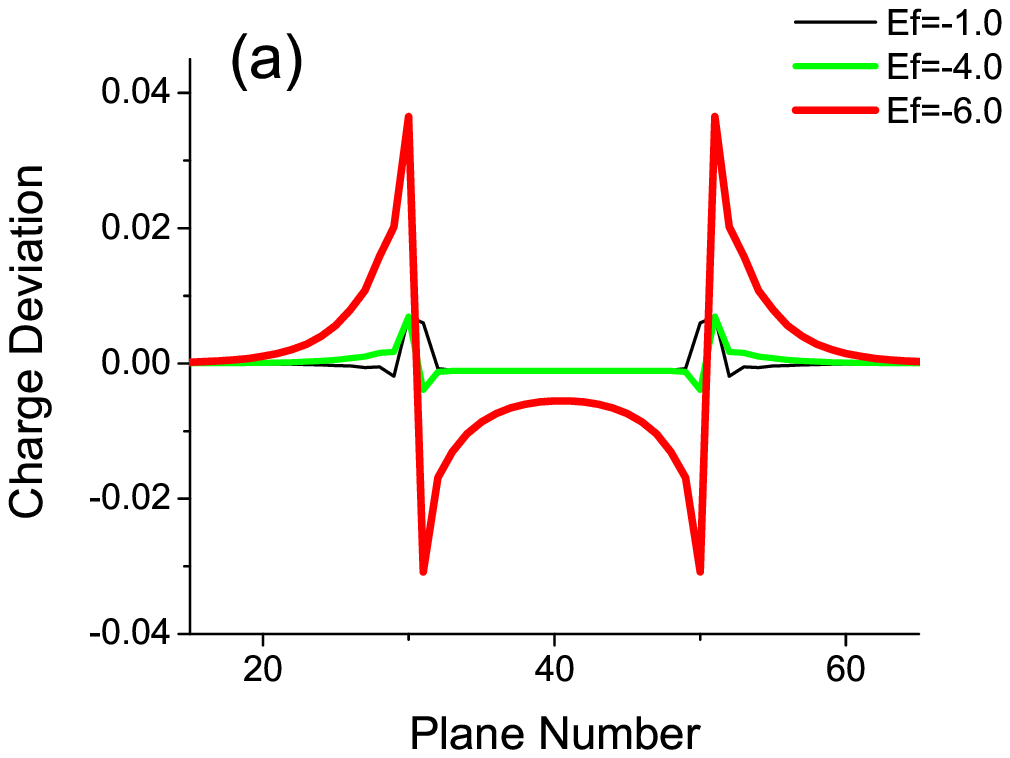}
\includegraphics[angle=0,width=8.5cm,clip=]{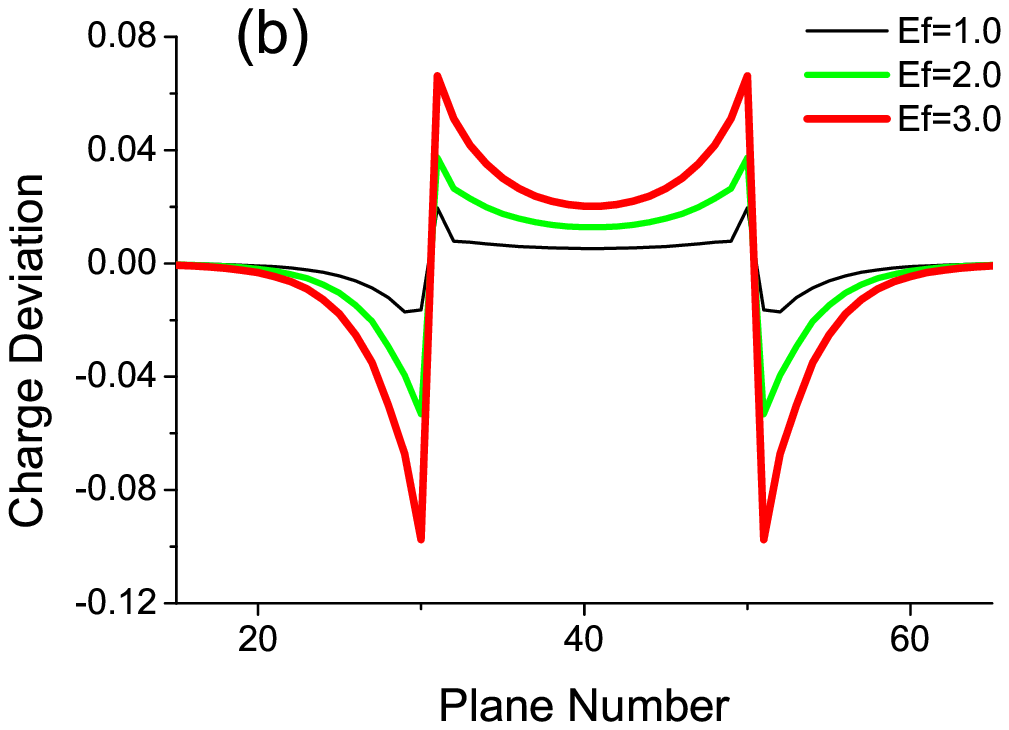}
\caption{Charge deviation in a multilayered nanostructure with a
barrier thickness of 20. The barrier is slightly doped with
electrons so that the conduction electron filling is 0.251
[$\mu^{bulk}_{barrier}(T=0.25)=7.046$]. The different lines denote
different $E_f$ values, as detailed in the legend. Note the charge
deviation pattern with (a) negative $E_f$ is different from the
pattern with (b) positive $E_f$. With positive $E_f$, the
magnitude of the charge deviation on the interface increases with
the magnitude of the shift. With negative $E_f$, if the magnitude
of the shifts are small, the magnitudes of the charge deviation do
not change too much due to the presence of the Mott gap. (Color on-line.)}
\label{mott0.251}
\end{figure}

The existence of a large band gap in the strongly correlated doped
insulator dramatically changes the charge reconstruction from the
metallic case. The device exhibits asymmetry with regard to
positive and negative band shifts as shown in
Fig.~\ref{mott0.251}. With positive band shifts, the carriers can
easily move from the leads to the barrier until the electric field
generated by electronic charge reconstruction strikes a balance
with the mismatch of the bands. Larger band shifts cause larger
electronic charge reconstruction at the interface. The situation
is more complicated for negative band shifts.

If the shift is small, the doped carriers are completely drained
from the barrier ($\rho_{e}\rightarrow 0.25$) but the band gap in
the Mott insulator stops further diffusion of the carriers into
the leads. The charge profile in this case is similar to that of a
Schottky barrier. In the n-type Schottky barrier, electrons in an
n-type semiconductor can lower their energy by traversing the
junction. As the electrons leave the semiconductor, a positive
charge, due to the ionized donor atoms, stays behind. This charge
creates a negative field and lowers the band edges of the
semiconductor. Electrons flow into the metal until equilibrium is
reached between the diffusion of electrons from the semiconductor
into the metal and the drift of electrons caused by the field
created by the ionized impurity atoms. The difference is that in a
Schottky barrier, a region in the semiconductor close to the
junction is depleted of mobile carriers while in our multilayer
nanostructure, the carriers are completely drained from the doped
barrier because of the small barrier thickness. (Note in
Fig.~\ref{mott0.251}(a), with small shifts, the charge deviations
in the barrier planes uniformly have the value of $-0.01$ except
for the planes close to the interfaces, which means all the doped
carriers move to the interfaces.) When the shifts get larger, the
carriers close to the interface begin to move to the leads. Thus
holes are created near the interface in the barrier side. When the
band shift gets large enough, the band gap can no longer stop the
diffusion of the carriers from the barrier to the leads. The
charge deviation curve at large negative band shifts bears a
similar shape to the curve at positive band shifts, although the
maximum net charges on the interface are significantly less due to
the effect of the Mott-insulating band gap.

\begin{figure}
\centering
\includegraphics[angle=0,width=8.5cm,clip=]{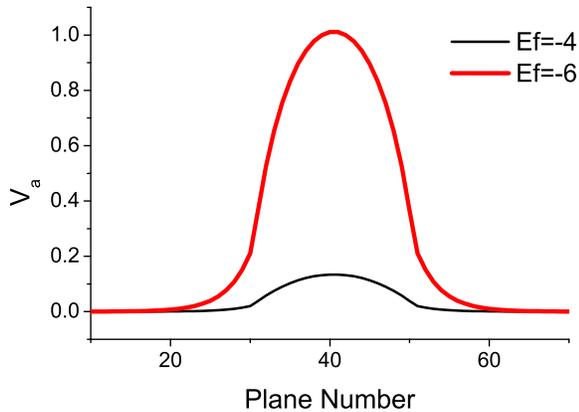}
\caption{Electric potential ($V_{\alpha}$) as a function of $E_f$.
The conduction electron filling is 0.251 and the interaction is
$U=12$. Note the drastic increase of the potential at the center
of the barrier from $E_{F}(T=0.25)=-4$ to $E_{F}(T=0.25)=-6$ (recall the bandgap
is equal to 5). (Color on-line.)} \label{potential}
\end{figure}

To better understand the effect of the band gap,
Fig.~\ref{potential} shows the electric potentials generated by
the charge reconstruction with $E_{F}=-4.0$ and $E_{F}=-6.0$
respectively. The nearly five fold increase of the potential at
the center of the barrier is because with $E_{F}=-6.0$, the band
gap can no longer hold the carriers in the lower band from moving
to the leads, resulting in more net charges along the interface.
This is quite different from the case where the barrier is
metallic, where an increase of $E_f$ generically results in a
smooth increase of the potential.

\begin{figure}
\centering
\includegraphics[angle=0,width=8.5cm,clip=]{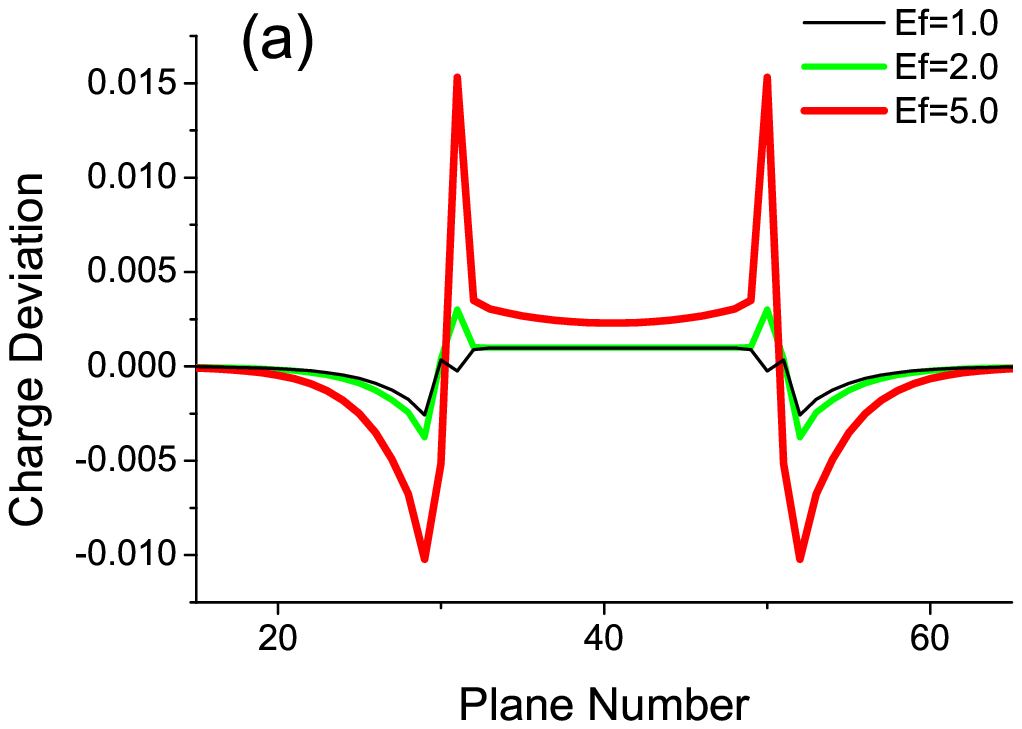}
\includegraphics[angle=0,width=8.5cm,clip=]{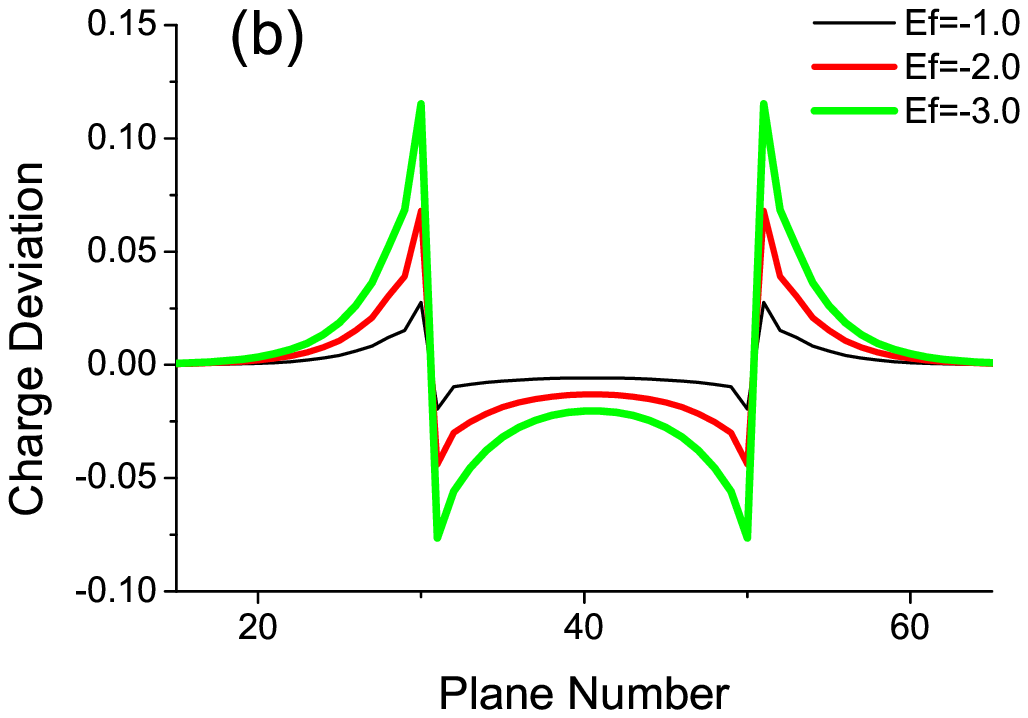}
\caption{Charge deviation in a multilayered nanostructure with a
barrier thickness of 20. The barrier is slightly doped with holes
so that the conduction electron filling is 0.249
[$\mu^{bulk}_{barrier}(T=0.25)=2.488$]. The different lines denote
different $E_f$ values, as detailed in the legend. The charge
deviation pattern in this case is the opposite of that with the
0.251 filling. With negative $E_f$ [panel(b)], the magnitude of the charge
deviation on the interface changes with the magnitude of the
shift. With positive $E_f$ [panel (a)], if the magnitude of the shift is
small, the magnitude of the charge deviation does not change
much. (Color on-line.)} \label{mott0.249}
\end{figure}
We find opposite results if the carriers are holes instead of
electrons. Fig.~\ref{mott0.249} shows the electronic charge
reconstruction for the doped Mott insulator with a conduction
electron charge density $\rho_e=0.249$. In this case, negative
band shifts attract holes from leads to the barrier. Larger shifts
create more net charges along the interface. Small positive band
shifts can only drain holes from the lower band of the barrier due
to the band gap. For large positive band shifts, the charge
deviation curve returns to the shape of those with negative band
shifts.
\begin{figure}
\centering
\includegraphics[angle=0,width=8.5cm,clip=]{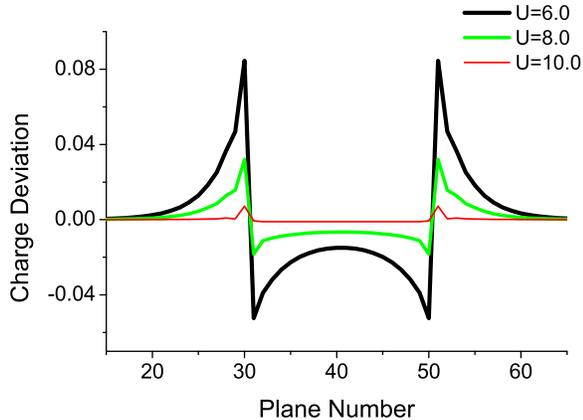}
\caption{Charge deviation as a function of interaction $U$. The
conduction electron filling is 0.251 and the shift $E_{f}$ is
fixed at $-2.0$ [$\mu^{bulk}_{barrier}(T=0.25)=1.792$ for $U=6$,
$\mu^{bulk}_{barrier}(T=0.25)=3.250$ for $U=8$ and
$\mu^{bulk}_{barrier}(T=0.25)=5.121$ for $U=10$]. Note the magnitude of
the charge deviation on the interface drops as $U$ increases.
However if $U$ is large enough, further increase of $U$ will not
result in a further decrease of the magnitude of the charge
deviation. (Color on-line.)} \label{change u}
\end{figure}

Next we examine the electronic charge reconstruction for a given
amount of band shift with different strengths of the interaction.
We choose electrons as the carriers ($\rho_e=0.251$). The band
shift $E_f$ is fixed at $-2.0$. Fig.~\ref{change u} shows the
different charge profiles as $U$ changes from $6$ to $10$. For
small $U$, the gap of the Mott insulator is small. The electrons
inside the barrier can easily diffuse into the leads. We see a
large electronic charge reconstruction at the interface. As $U$
gets larger, it becomes harder for the electrons to move into the
leads due to the larger band gap. The net amount of the charge
deviation at the interface becomes less. With strong enough
interaction, only the doped carriers can be drawn from the barrier.
The curve then resembles the one with $U=12$.

In our model, the screening length $e_{Schot}$ can be changed in
the leads as well as in the barrier. We tried several different
values of $e_{Schot}$ ranging from 0.1 to 2.0. With a larger
screening length, the magnitude of the charge reconstruction increases,
essentially because the electric field generated by the
charge reconstruction is stronger. There are more net charges on
the planes close to the interface. Planes deeper inside the leads
are less affected by the change in the screening length. The effect of
$e_{Schot}$ is less obvious in the doped Mott insulator case with
large interaction.  The total number of carriers is limited in this
case. A longer screening length cannot move more carriers into the
leads, because they are depleted near the interface and the field is not
strong enough to start moving charge out of the lower Hubbard band.

The thickness of the barrier is not playing an important role in
these multi-layered structures (for moderately thick barriers). 
We adjusted the thickness of the
barrier from 10 to 20 planes. More planes in the barrier obviously
provides more carriers, generating more net charges on the
interface. But the properties of the central part of the barrier
remain similar.

\section{\normalsize{CONCLUSION}}
In this paper, we applied a generalized DMFT to inhomogeneous
systems to calculate the self-consistent many-body solutions for
multilayered nanostructures with barriers that can be adjusted to
go through the Mott transition. We developed the computational
formalism based on the algorithm of Potthoff and Nolting and the
Falicov-Kimball model that can calculate the charge reconstruction
of the multilayered nanostructures with variable barrier fillings.
We focused our study on the doped Mott insulator as the barrier
material. We found interesting results that came out of this
analysis.

First of all, the scaling effect (the charge reconstruction
divided by the value of $E_f$ remain roughly the same across the
system) at half filling in the metal phase is no longer valid at
fillings other than 0.5. Second, the symmetry of the charge
reconstruction with positive and negative $E_f$ in the metal phase
is broken in the Mott insulating phase. With electrons as carriers
in the doped barrier, small negative $E_f$ can only draw the doped
carrier to the leads. If the value of $|E_{f}|$ is smaller than
the size of the band gap of the Mott insulator, the net charges on
the interface do not increase much with the increase of $|E_{f}|$
and the central part of the barrier remains drained of carriers. If
the the value of $|E_{f}|$ rises above the size of the band gap of
the Mott insulator, electrons from the lower band start to move to
the leads. The charge profile changes drastically. With holes as
carriers in the doped barrier, the positive region of $E_f$ is
affected by the band gap. So the charge reconstruction is opposite
to the electron doped barrier with respect to $E_f$. Third, the
charge reconstruction of the doped Mott insulator case does not
change much with changing parameters like $e_{Schot}$ or the
thickness of the barrier. The central planes of the barriers are
always drained of carriers. And the amount of the net charges on
the interface is largely determined by the number of carriers
inside the barrier with the strength of the interaction smaller
than the band gap. That means physical properties of devices made
in this structure may be stable even when the parameters of the
barrier material mentioned above have some deviation. So the next
step is to calculate the transport properties or the resistance of
this multilayered nanostructure with the doped Mott insulating
barriers and to compare the results from different parameter sets.
This requires generalizing the codes to the real axis, which is
beyond the scope of this work.
With small band shifts, we anticipate that the resistance should not change much
because the charge reconstruction is dominated by the amount of
carriers in the barrier. Because the carriers in the barrier are
completely drained, the resistance should be large. With a large
band shift, the resistance should be significantly lower since
charges from the lower band now move to the lead, and the whole
structure is more conductive as a result (even though the Coulomb potentials
are larger).

\end{document}